\documentclass[aps,prl,10pt,notitlepage,twocolumn,superscriptaddress,raggedbottom]{revtex4-1}
\usepackage{graphicx}
\usepackage{amsmath}
\usepackage{amssymb}
\usepackage{color}
\usepackage{amscd}
\usepackage{bm}
\usepackage{enumerate}

\usepackage{mathrsfs}
\usepackage{calrsfs}
\usepackage{epsfig}
\usepackage{subfigure}
\usepackage{psfrag}

\def\beq{\begin{equation}}
\def\eeq{\end{equation}}
\def\bea{\begin{eqnarray}}
\def\eea{\end{eqnarray}}

\begin{document}
\title{Propagating stress waves during epithelial expansion}
\author{Shiladitya Banerjee}
\thanks{To whom correspondence may be addressed. Email: shiladityab@uchicago.edu or mcmarche@syr.edu}
\affiliation{James Franck Institute, The University of Chicago, Chicago IL 60637}
\author{Kazage J.C. Utuje}
\affiliation{Department of Physics, Syracuse University, Syracuse NY 13244}
\author{M. Cristina Marchetti}
\thanks{To whom correspondence may be addressed. Email: shiladityab@uchicago.edu or mcmarche@syr.edu}
\affiliation{Department of Physics, Syracuse University, Syracuse NY 13244}
\affiliation{Syracuse Biomaterials Institute, Syracuse University, Syracuse NY 13244}

%\date{\today}

\begin{abstract}
Coordinated motion of cell monolayers during epithelial wound healing and tissue morphogenesis involves mechanical stress generation. Here we propose a model for the dynamics of epithelial expansion that couples mechanical deformations in the tissue to contractile activity and polarization in the cells. A new ingredient of our model is a feedback between local strain, polarization and contractility that naturally yields a mechanism for viscoelasticity and effective inertia in the cell monolayer. Using a combination of analytical and numerical techniques, we demonstrate that our model quantitatively reproduces many experimental findings [Nat. Phys. {\bf 8}, 628 (2012)], including the build-up of intercellular stresses, and the existence of traveling mechanical waves guiding the oscillatory monolayer expansion. \\ 


\end{abstract}

\maketitle
Many developmental processes, such as embryogenesis~\cite{mammoto2010}, tissue morphogenesis~\cite{keller2002}, wound healing~\cite{fenteany2000} and cancer metastasis~\cite{tsuji2009},  involve collective cell migration~\cite{weijer2009} and long-scale force generation, which in turn rely on the interplay of cell-cell cohesion, cell adhesion to the extracellular matrix, as well as myosin based contractility~\cite{manning2010,mertz2013}. Recent experiments reveal that unconstrained tissue expansion is accompanied by propagating mechanical waves and build-up of intercellular stresses~\cite{serra2012}. These waves are controlled by expressions of myosin activity, cell-cell adhesion and cytoskeletal remodeling. These findings pose a fundamental physical question: how do waves arise in over-damped active elastic media? What are the underlying spatio-temporal patterns governing stress propagation in dense expanding cell layers?

Active materials encompass a wide range of living and non-living systems with inborn mechanical stresses regulated by chemical reactions. Generic descriptions of the dynamics of such materials predict a broad class of non-equilibrium states including spontaneous flow, wave propagation and pattern formation~\cite{kruse2005,bois2011,marchetti2013,kumar2014}. While the dynamics of active fluids have been extensively studied, quantitative descriptions of active contractile materials are much less developed. Recent work has suggested that a polarized elastic medium driven by chemical agents can exhibit finger-like protrusions and internal stress accumulation during expansion~\cite{arciero2011,kopf2013}.  
It remains unclear, however, how cell contractility, polarization or tissue cohesion influence stress generation and wave propagation. Earlier work by two of us and others showed that the coupling of mechanical and chemical degrees of freedom can lead to an effective inertia and sustained propagation of waves~\cite{banerjee2011,banerjee2011a,radszuweit2013}. Related models also emphasize that turnovers in actomyosin activity are essential to capture spontaneous oscillations in cell cytoskeleton~\cite{placcais2009,dierkes2014}. In this Letter, we propose a new mechanism of stress propagation in multicellular materials based on a local feedback between elastic deformations and cell contractility.

We consider a minimal model for an expanding cell monolayer, described as an elastic continuum coupled to an internal degree of freedom, the concentration of active contractile units. The assumption of elasticity is supported by experimental evidence that in cohesive cell layers stress and strain tend to be in phase, as in elastic materials~\cite{serra2012,mertz2013}. The contractile units represent actomyosin assemblies that locally generate contractile stresses in the cells. We propose that tissue expansion promotes the rate of assembly of these contractile units, leading to larger contractile forces that can compete with propulsion forces. This  mechano-chemical feedback successfully captures the experimentally observed stress waves~\cite{serra2012}. The steady state of such a system is described by polarization being largest at the edges and lowest at the center. 
%In addition, we also consider the effect of time-dependent propulsion forces due to fluctuations in cell polarization. Finally, 
A scaling model for the expanding cell layer captures the mechanical oscillations and predicts self-sustained periods of stiffening and fluidization in the tissue.

\begin{figure}
\centering
\includegraphics[width=\columnwidth]{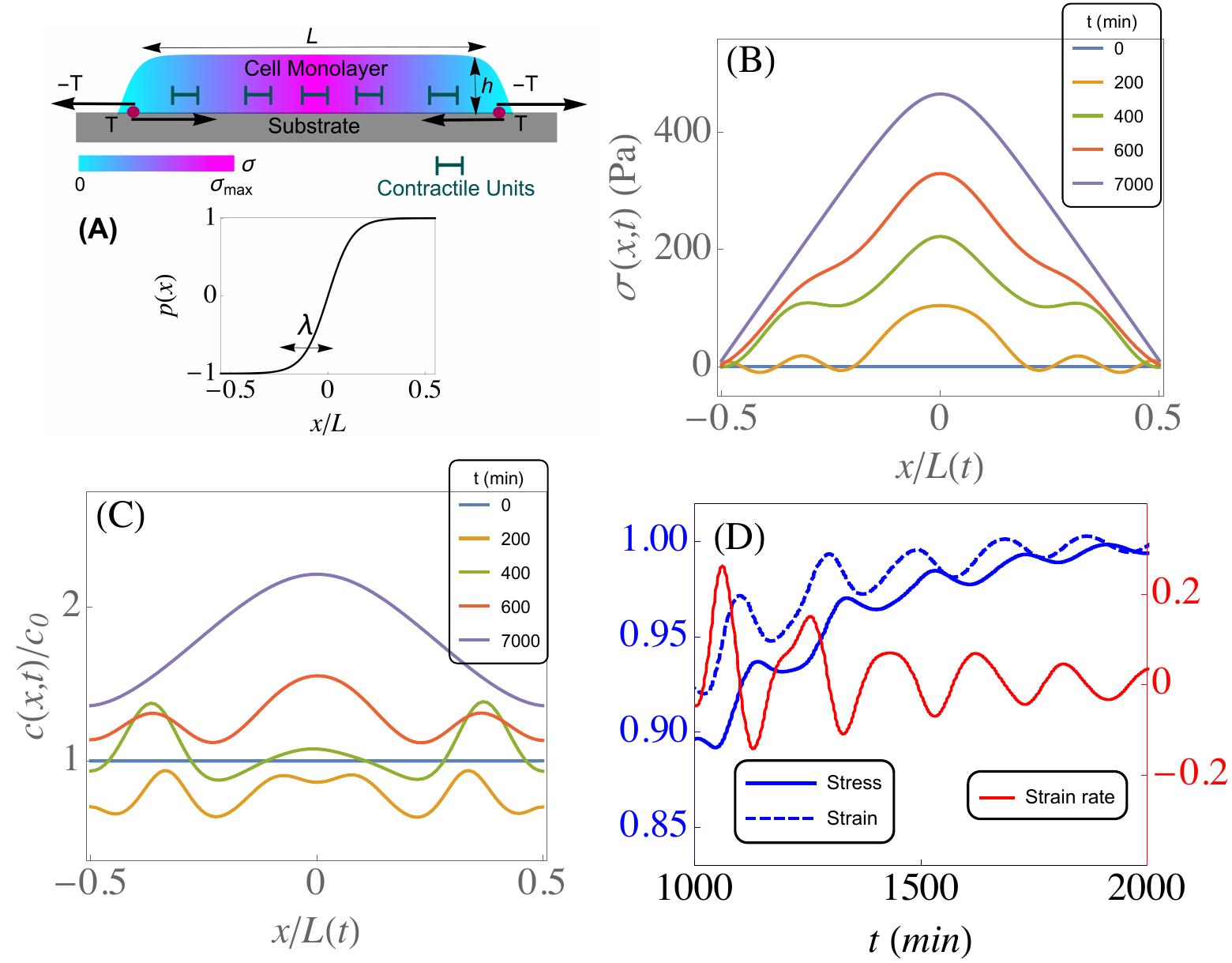}
\caption{(color online) (A) Top: Schematic of a spreading cell monolayer. Traction stresses (${\bf T}$) are indicated by arrows and the colormap denotes local magnitude of monolayer stress. Bottom: Profile of cell polarization. (B) Time-evolution of the internal stress $\sigma(x,t)$ in the monolayer. (C) Time-evolution of the concentration of contractile units, $c$, normalized by its equilibrium value. (D) Midline stress $\sigma(0,t)/\sigma_\infty(0,t)$ (blue solid), midline strain $\varepsilon(0,t)/\varepsilon_\infty(0,t)$ (blue dashed) and midline strain rate $\dot{\varepsilon}(0,t)$ (red solid, units $10^{-4}$ s$^{-1}$) as functions of time. Parameters: $B=120$ Pa, $\beta=200$ Pa, $\Pi/\beta=10^{-3}$, $\tau=350$ min, $\alpha/c_0=1/560$ min$^{-1}$, $L_0=600$ $\mu$m, $h_0=6$ $\mu$m, $f_0=4$ Pa, $\lambda=30\ \mu$m, $\Gamma=0.009$ nN min/$\mu$m$^3$, $D=26$ $\mu$m$^2$/min. }
\label{fig:stress}
\end{figure}

\paragraph{Continuum Model for spreading cell layer.}
We consider a thin film of cell monolayer spreading in the $x$-$y$ plane, with height $h(t)$ and length $L(t)$ at time $t$ (Fig.~\ref{fig:stress}A, inset). In the absence of external forces, in-plane force-balance gives $\partial_j\Sigma_{ij}+\partial_z\Sigma_{iz}=0$, where ${\bm \Sigma}$ is the stress tensor and the latin indices denote in-plane coordinates $x,y$. For $h\ll$ $L,d$, the $x$ and $y$ linear extensions of the cell layer, we average the force-balance equation across the z-direction to obtain $h\partial_j \sigma_{ij}=\Sigma_{iz}\vert_{z=0}$, where ${\bm \sigma}(x,y)=\frac{1}{h}\int_0^h dz{\bm \Sigma} (x,y,z)$, assuming that the top layer ($z=h$) is stress free. The shear stress at the cell-substrate interface is the traction stress exerted by the cell on the substrate. It is given by, $\Sigma_{iz}\vert_{z=0}=T_i=\Gamma \partial_t u_i - f_0 p_i$, with $\Gamma$ the friction density, ${\bf u}$ the elastic displacement field, ${\bf p}$  the cell polarization and $f_0$ the propulsion force per unit cross-sectional area. The term $f_0p_i$ is supported by the experimental observation that the local velocity of expanding monolayers is generally not aligned with traction, requiring the existence of an internally generated driving force associated with cell motility~\cite{kim2013}. Both $\Gamma$ and $f_0$ are controlled by integrin-mediated cell-environment interactions.
%By force balance the substrate  exerts a thrust force $-T_i$ on the cell.  
We further simplify the model by assuming translational invariance along the y-direction. The equation of motion governing the displacement field, $u(x,t)$, of the cell layer is ($0<\vert x\vert<L/2$),
\begin{equation}\label{eq:u}
\Gamma \partial_t u = f_0 p(x,t) + h(t) \partial_x \sigma\;,
\end{equation}
where $\sigma(x,t)$ is the internal stress in the monolayer, $\sigma=-\Pi + B \varepsilon + \sigma_a (c)$. It is given by the sum of an internal pressure ($\Pi$), an elastic stress, with $B$ the compressional elastic modulus and $\varepsilon=\partial_x u$ the strain field,  and an active stress $\sigma_a$ that depends on the concentration $c$ of active contractile units, such as phosphorylated myosins interacting with actin filaments. The constant pressure $\Pi$ accounts for internal growth due to cell proliferation which is assumed negligible without loss of generality. The active stress is proportional to the chemical potential of the active species, $\mu$, which we take proportional to the logarithm of the concentration of the species.
%which is related to the concentration field as, $\Delta\mu(c)\propto \log{(c/c_0)}$, 
%with $c_0$ the equilibrium concentration~\cite{SMprl14}. 
We thus have $\sigma_\text{a}(c)=\beta \log{(c/c_0)}$, where $c_0$ is the concentration of contractile elements in equilibrium ($f_0=0$) and $\beta>0$ the magnitude of the contractile stress. The dynamics of the concentration field $c(x,t)$ is given by,
\begin{equation}\label{eq:c}
\partial_t c = -\frac{1}{\tau}(c-c_0)  + \alpha\varepsilon- \partial_x J\;,
\end{equation}
where $\tau$ is the timescale of turnover of the contractile elements, $\alpha>0$  is the rate of production  of $c$ due to local extension (or degradation due to contraction) and $J (x,t)$ is the current responsible for transport of these active units. This is in contrast to our earlier works~\cite{banerjee2011,banerjee2011a}, where the strain field enters the dynamics of $c$ through the decay rate. The total current is a sum of diffusive and convective fluxes, $J=-D\partial_x c + c\partial_t u $, where $D$ is an effective diffusion constant, describing the tendency of neighboring cells to equalize activity levels. Together Eqs.~\eqref{eq:u} and \eqref{eq:c} define the dynamics of the spreading monolayer, given the form of $p(x,t)$, the boundary and initial conditions. We first consider the case of constant but non-uniform propulsion force given by $p(x,t)=\tanh{(x/\lambda)}$ where $\lambda$ is a length scale controlling the width of the transition zone from left moving to right moving cells at the center of the monolayer (see Fig.~\ref{fig:stress}A). The length of the spreading layer at time $t$ is given by, $L(t)=L_0 + u(L_0/2,t)-u(-L_0/2,t)$, and the height is determined by the condition of volume conservation, $h(t)=h_0L_0/L(t)$, with $L_0$ and $h_0$  the initial length and height of the monolayer, respectively. The boundary of the monolayer is stress free, i.e., $\sigma(\pm L/2,t)=0$ at all times. 
We assume that the monolayer is initially undeformed, $u(x,0)=0$, with an equilibrium concentration of contractile elements, $c(x,0)=c_0$, and choose a no-flux boundary condition for $c$, $\partial_x c(\pm L/2,t)=0$.

\begin{figure}
\centering
\includegraphics[width=\columnwidth]{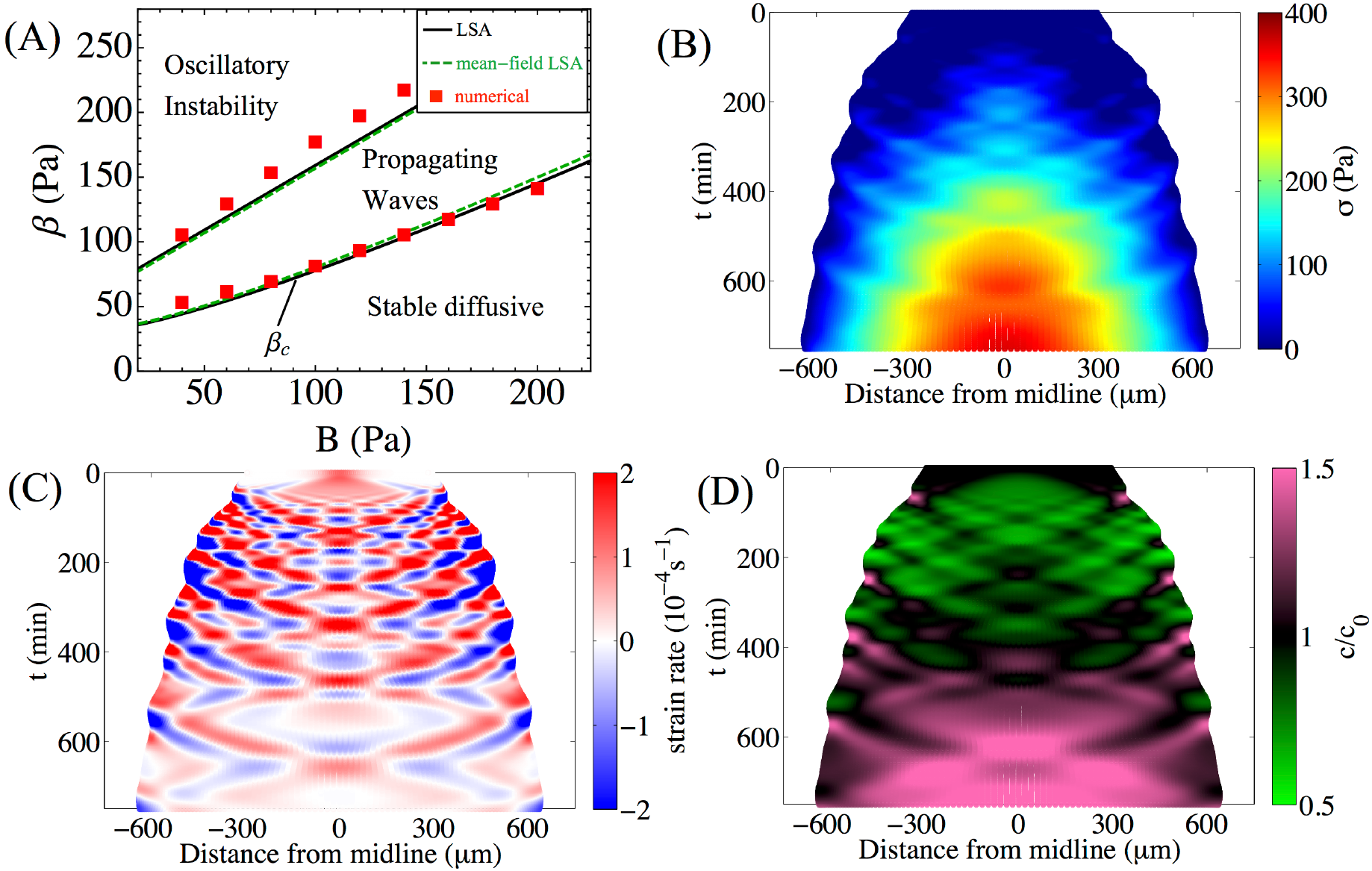}
\caption{(color online) (A) Phase diagram of the spreading gel. The vertical axis represents the contractile activity $\beta$ and the horizontal axis is the compressional modulus $B$. Three behaviors are observed: stable diffusive,  stable propagating waves, and  oscillatory instability. The red squares are obtained from the numerical solutions of the full nonlinear model, the black solid lines are the results of the linear stability analysis (LSA) of the equilibrium state (at $q=13.5/L_0$)~\cite{SMprl14}, and the dashed green lines refer to the LSA of the mean-field model given in Eqs.~\eqref{MF}. Kymographs of (B) the monolayer stress field, (C) strain rate $\partial_t\varepsilon(x,t)$, and (D) $c(x,t)/c_0$. The parameter values are taken to be the same as in Fig.~\ref{fig:stress}.}
\label{fig:waves}
\end{figure}

\paragraph{Propagating waves.}
In the absence of propulsion force ($f_0=0$), the cell layer is in a quiescent homogeneous state, with $u=0$ and $c=c_0$. When $f_0\not=0$, the cell layer spreads and reaches a steady-state at long times.
%with stress  $\sigma_{\infty}(x)=-(f_0\lambda/h_{\infty})\log{\left(\frac{\cosh{(x/\lambda)}}{\cosh{(L_\infty/2\lambda)}}\right)}$ for $\lambda\ll L$, with   $L_\infty$ and  $h_\infty$ the asymptotic length and height (see SI).} 
We have integrated numerically  Eqs.~(\ref{eq:u},\ref{eq:c}) with the given initial and boundary conditions, using the Runge-Kutta-Fehlberg method. The model parameters are chosen to quantitatively describe the available experimental data for MDCK colonies~\cite{serra2012,SMprl14}.
The phase diagram shown in Fig.~\ref{fig:waves}A displays  three dynamical regimes in terms of  contractile activity $\beta$ and compressional modulus $B$ (controlled by cell-cell adhesion): a region where fluctuations are stable and diffusive  at low contractility, an intermediate region where the system supports propagating waves, and a region where the propagating waves become unstable at high contractility. There is good agreement between the boundaries obtained via numerical solution of the full nonlinear equations (red diamonds) and those determined by the linear instability of fluctuations about the equilibrium, undeformed state~\cite{SMprl14} and about the long-time solution of the mean-field model in Eqs.\eqref{MF}. 
In the region of propagating waves, the stress initially shows a few local maxima (Fig.~\ref{fig:stress}B),  which evolve towards a single maximum at the center of the monolayer, as observed in  experiments~\cite{trepat2009,serra2012}. The concentration of contractile elements  also  oscillates and builds up at the center of the monolayer (Fig.~\ref{fig:stress}C). The stress waves propagate nearly in phase with the strain field, whereas the strain rate fluctuates nearly out of phase with the stress (Fig.~\ref{fig:stress}D). Thus the response of the material is dominated by elastic relaxation with dissipation induced by turnovers in contractility on a timescale $\tau$. The waves span the entire length of the monolayer and consist of a strain rate wavefront that propagates inwards from the edge, and then travels back to the edge, resembling an X-pattern, as observed experimentally~\cite{serra2012}. 
With the given parameter values our numerical simulations capture the mechanical waves as evident in the kymographs of stress, strain rate and concentration of contractile units (Fig.~\ref{fig:waves}B-D).  

\begin{figure}
\centering
\includegraphics[width=\columnwidth]{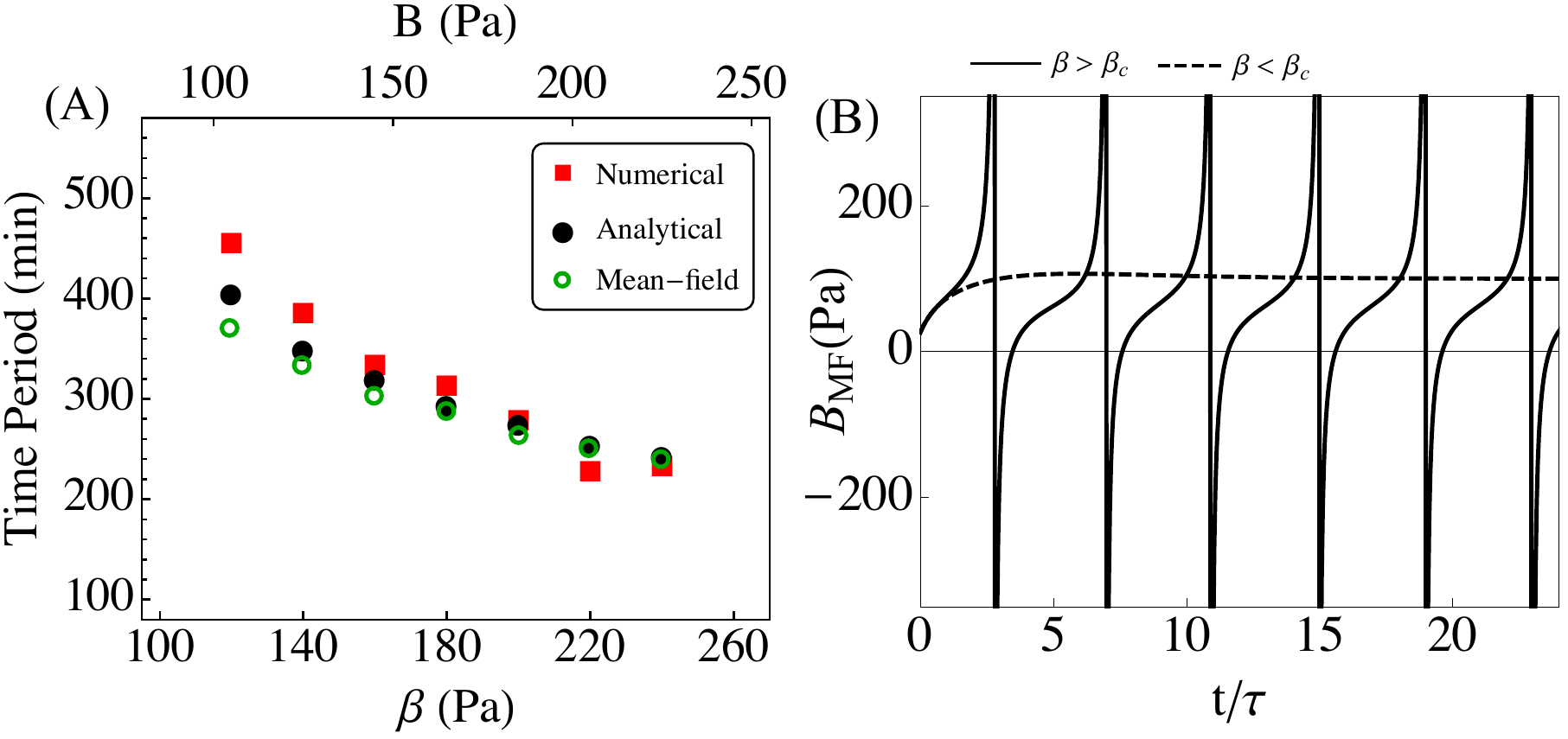}
\caption{(color online) (A) Period of oscillation determined from the numerical solution to Eqs.~(\ref{eq:u},\ref{eq:c}) (red squares), obtained from Eq.~\eqref{eq:strain} (black solid circles), and as predicted by the mean-field model (green open circles) for various  $\beta$ and $B$. (B) Mean-filed elastic modulus $B_\text{MF}$ of the cell monolayer as a function of time, showing oscillatory stiffening/fluidization for $\beta=100$ Pa (solid) and steady stiffening for $\beta=30$ Pa (dashed). Parameters: $B=60$ Pa, $\tau=350$ min, $c_0/\alpha=780$ min, $F_0=8$ nN, $\gamma=9$ nN min/$\mu$m, $dh_0/L_0^2=0.1$.}
\label{fig:meanfield}
\end{figure}

To understand the origin of wave propagation and estimate the wave frequency, it is useful to examine the linear fluctuations in the strain field, $\delta \varepsilon$ and the concentration field $\delta c$, about the quiescent homogeneous state, $u=0$, $c=c_0$ and no spreading force. Using Eqs.~\eqref{eq:u} and \eqref{eq:c}, one can then eliminate $\delta c$ from such linearized equations to obtain the linearized dynamics of strain fluctuations,
\begin{equation}\label{eq:strain}
\tau\Gamma\partial_t^2 \delta\varepsilon + \Gamma\partial_t \delta\varepsilon = h_0\left(B_\text{eff}+ \eta_\text{eff}\partial_t - \tau BD \partial_x^2\right)\partial_x^2\delta\varepsilon  \;,
\end{equation}
%where $f(x)=f_0c_0[p'(x)\tau^{-1}-Dp'''(x)]/\beta$, is the internally generated driving force associated with strain growth.}
The above equation shows that the coupling of strain to concentration field yields an effective mass density (inertia), $\tau\Gamma$, and viscoelasticity characterized by an effective elastic modulus, $B_\text{eff}=B+\alpha\beta\tau/c_0$, and an effective viscosity $\eta_\text{eff}=(B-\beta + D\Gamma/h_0)\tau$. The dynamics of strain fluctuations resembles a damped Kelvin-Voigt oscillator with a characteristic frequency of oscillations, $\omega_0=q\sqrt{h_0\left(B_\text{eff}+\tau q^2BD\right)/(\tau\Gamma)}$, with $q$ the wavevector. The estimate for the time period $2\pi/\omega_0$ agrees well with the time period determined from numerics for $q\simeq 4\pi/L_0$ (see Fig.~\ref{fig:meanfield}A) and with the value measured in recent experiments~\cite{serra2012}. Finally, we note that if the concentration $c$ is conserved ($\tau\rightarrow \infty$; $\alpha=0$), stable propagating waves are spontaneously generated for $0<B-\beta+D\Gamma/h_0<2\sqrt{DB\Gamma/h_0}$. If diffusion is slow compared to elastic relaxation, $D\Gamma/Bh_0\ll 1$, stable propagating waves are not observed~\cite{SMprl14}. In the opposite limit of infinitely fast turnovers in contractility ($\tau \rightarrow 0$), strain fluctuations decay diffusively at a rate $\simeq Bh_0/\Gamma L^2$.

%For $\alpha>0$ and a fixed value of elastic modulus $B$, the phase diagram in Fig.~\ref{fig:waves}A shows that the propagating waves occur at finite values of the wavenumber and are not observed in the long wavelength limit ($q\rightarrow 0$).
%The oscillations are unstable when the effective damping coefficient in Eq.~\eqref{eq:strain}, $\Gamma_\text{eff}\simeq \Gamma + \eta_\text{eff}/L^2$, changes sign for $\beta>\Gamma L^2/\tau + B+D\Gamma/h_0$. 
%If $\alpha<0$, the waves disappear and there exists a long wavelength {\it contractile} instability for $\beta>Bc_0/\alpha\tau$, describing material failure~\cite{banerjee2011,wang2013}. {\color{red}In the absence of contractility ($\beta=0$) the long-wavelength instabilities do not arise for $\alpha<0$, as discussed in Ref.~\cite{kopf2013b}}.

%A linear stability analysis of the inhomogeneous stationary solution $(\varepsilon_\infty(x),c_\infty(x))$ yields the range of parameters where propagating waves can occur. 

\paragraph{Mean field model.}
The mean field limit of the continuum model is obtained by neglecting spatial variations in  $c$ and  $\varepsilon$ and it is formulated in terms of the length ($L$), height ($h$), and the average concentration of contractile elements, $\overline{c}(t)=\frac{1}{L}\int_0^L dx\ c(x,t)$, with
\begin{subequations}
\label{MF}
\begin{gather}
\gamma \frac{dL}{dt}=F_0 - A(t) \sigma(t)\;,\\
\frac{d\overline{c}}{dt} + \frac{\overline{c}}{L}\frac{dL}{dt}=-\frac{1}{\tau} (\overline{c}-c_0) + \alpha \varepsilon\;,
\end{gather} 
\end{subequations}
with $F_0$ the propulsion force, $\gamma$ the friction, $A(t)=dh(t)$ the cross-sectional area, $\varepsilon(t)=L(t)/L_0-1$ the strain and $\sigma(t)$ the internal stress given by $\sigma(t)=B\varepsilon(t) + \beta(\overline{c}(t)/c_0-1)$. The height is determined using the incompressibility condition, with the size in the $y$ direction, $d$, fixed. The steady state solution is $L_\infty=L_0/(1-\Lambda)$, $h_\infty=h_0(1-\Lambda)$ and $c_\infty=c_0+\alpha\tau\Lambda/(1-\Lambda)$, with $\Lambda=c_0F_0/dh_0(Bc_0 + \alpha\beta\tau)$ the net compressive strain in the $z$-direction. For a given value of elastic modulus $B$, the mean-field model predicts oscillatory solutions for $\beta>\beta_c$, where $\beta_c(B)$ defines the phase boundary in $(B, \beta)$ plane separating the regions of propagating waves and diffusive spreading (dashed line in Fig.~\ref{fig:waves}A).
%The stress waves manifest as shape oscillations in the growing length and the shrinking height of the cell monolayer \mcm{(Fig.~\ref{fig:meanfield}A).}
For $\beta<\beta_c$ the monolayer diffusively approaches the steady state $(c_\infty, L_\infty)$.
%Dynamics of length (red), height (green) and concentration (blue) normalized by their initial values for $\beta=30~{\rm Pa}<\beta_c$ (top) and $\beta=100~{\rm Pa}>\beta_c$ (bottom).
%The numerical value for the time period is obtained by fast fourier transform of the strain rate at the center of the monolayer
This simple mean-field approach allows us to study the  material response of the monolayer characterized by an effective elastic modulus, $B_\text{MF}=d\sigma/d\varepsilon$. The oscillatory regime ($\beta>\beta_c$) exhibits sustained oscillations in the material rigidity, $B_\text{MF}$, with a slow period of stiffening followed by a sharp turnover (see Fig.~\ref{fig:meanfield}B). For $\beta<\beta_c$, the material gradually stiffens with $B_\text{MF}$ asymptotically approaching the value $B_\text{eff}$.
These oscillations reflect self-sustained turnovers in the cytoskeleton with periodic reinforcement and fluidization on different timescales, which was invoked to be the underlying mechanism of wave propagation in Ref.~\cite{serra2012}.

\paragraph{Time-dependent propulsion forces.}
Finally, we consider time variations of the propulsion force, as arising from the dynamics of cell polarization $p(x,t)$ given by
\begin{equation}\label{eq:p}
\partial_t p= (a-b p^2) p + \kappa \partial_x^2 p - w\partial_x\varepsilon + w'\partial_x(c/c_0)\;,
\end{equation}
where the first two terms with $b>0$ allow for the onset of a homogeneous polarized state when $a>0$. The stiffness constant $\kappa$ characterizes the cost of local deformations in the polarization. The last two terms in Eq.~\eqref{eq:p} define  active couplings of $p$ to the strain and the concentration field, with $w, w'>0$, such that $p$ aligns with the gradient of monolayer density and the concentration field. In other words, cell polarization is enhanced in the direction opposite to that of elastic restoring forces. Additionally, polarization gradients can induce mechanical stresses, and the stress tensor is modified to read, $\sigma=B\varepsilon + \sigma_a(c) + \beta' \partial_x p$, where $\beta'>0$ is a contractile tension induced by polarization gradients. We assume a no-flux boundary condition, $p'(\pm L/2)=0$. For $w=w'=0$ and if $t\gg a^{-1}$, such that $L\gg\sqrt{\kappa/a}$, the solution is essentially time-independent, and can be approximated as, $p_\infty(x)\simeq\sqrt{(a/b)}\tanh{(x/\lambda)}$, with $\lambda=\sqrt{\kappa/a}$.

\begin{figure}
\centering
\includegraphics[width=\columnwidth]{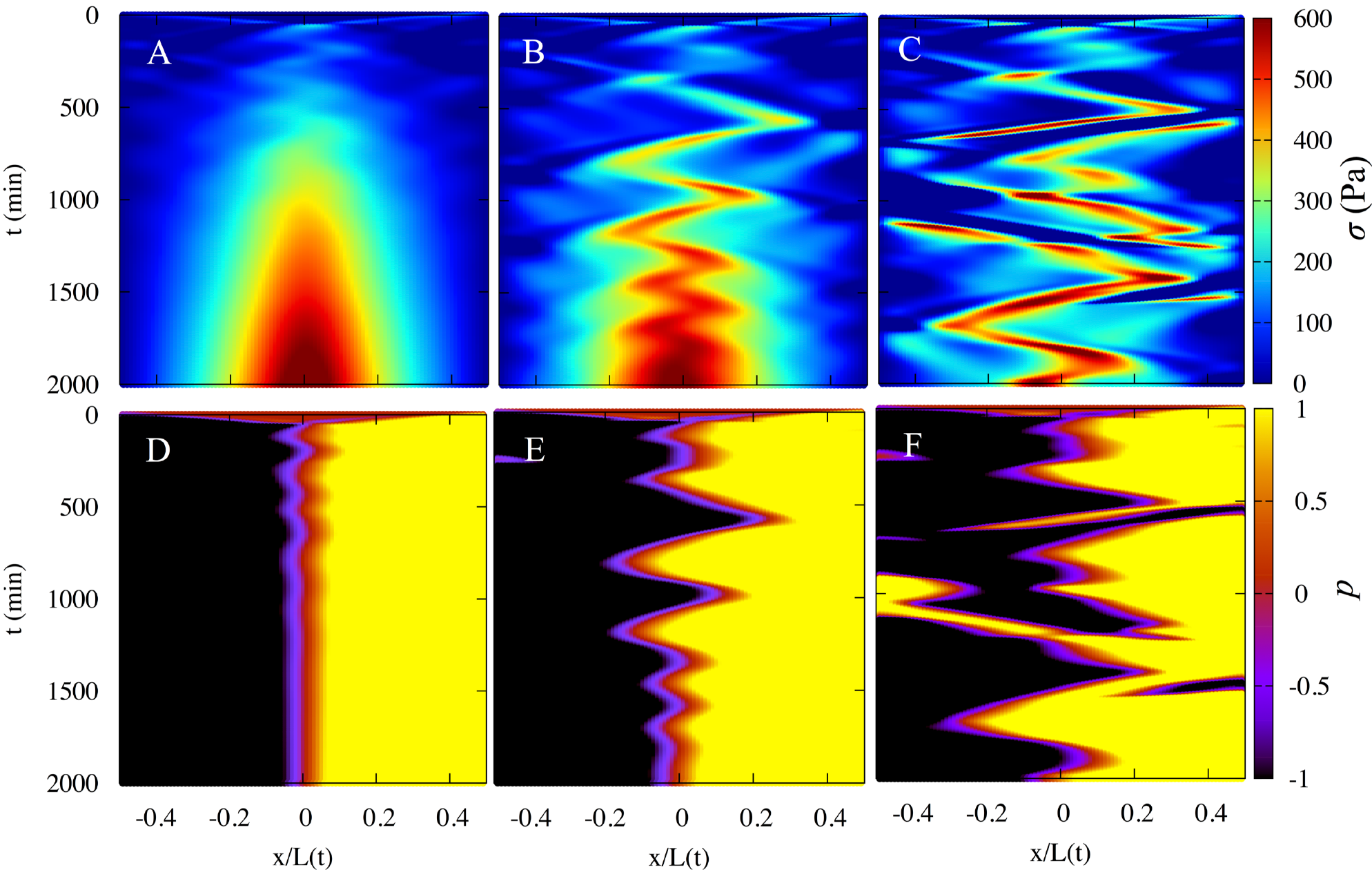}
\caption{(color online) Spatio-temporal evolution of internal stress (A-C) and polarization (D-F) as the polarization induced tension $\beta'$ is increased (left to right). (A,D) X-waves, $\beta'=12$ nN/$\mu$m; (B,E) traveling stress pulse, $\beta'=17$ nN/$\mu$m; (C,F) complex oscillatory patterns, $\beta'=24$ nN/$\mu$m. Parameters: $w=4.3$ $\mu$m/min, $w'=0.21$ $\mu$m/min, $\kappa=193$ $\mu$m$^2$/min, $a=0.07$ min$^{-1}$, $b=0.03$ min$^{-1}$. Other parameter values are the same as in Fig.~\ref{fig:stress}. See Supplemental Material~\cite{SMprl14} for kymographs of strain rate, velocity and the traction stress.}
\label{fig:pol}
\end{figure}

When the coupling of polarization to strain and contractility is turned on, various spatiotemporal patterns emerge as the active tension $\beta'$ is varied. For small $\beta'$, the stress patterns are qualitatively similar to Fig.~\ref{fig:waves}B (with time-independent propulsion), and $p$ asymptotically approaches $p_\infty$ with initial oscillations near the midline (Fig.~\ref{fig:pol} A,D). For intermediate  $\beta'$, a traveling stress pulse emerges in the layer and the location of stress maxima oscillate around the midline (Fig.~\ref{fig:pol}B). This is accompanied by large amplitude oscillations of net polarity that attenuate in time to generate a symmetric steady state polarization profile (Fig.~\ref{fig:pol}E). These traveling pulses persist even in the case $\beta=0$. For even higher values of $\beta'$ complex oscillatory patterns emerge in the monolayer stress and polarization (Fig.~\ref{fig:pol}C,F).

\paragraph{Discussions.}
We have developed a simple yet rich dynamic model for an active spreading gel, based on a linear feedback between local strain and contractility. A local increase in length due to spreading promotes the assembly of active elements that in turn induce contraction. We propose that a finite turnover rate in the active contractile elements can yield an effective inertia and viscoelasticity in the gel that vanishes for infinitely fast turnover rates.
This simple mechano-chemical model allows us to capture the experimentally observed propagating stress waves during tissue expansion without invoking nonlinear elasticity~\cite{serra2012}. These stress waves are characterized by strain rate wavefronts that initiate from the leading edge and periodically travel into and away from the midline of the monolayer. Our findings also elucidate that the effective material rigidity of the tissue undergoes sustained periods of stiffening and softening as the waves propagate. 
%We draw two important conclusions about the mechanical coordination in cell monolayers. %Secondly, the existence of mechanical waves depends upon a local feedback between material strain and contractility {\color{red}and does not require nonlinear elasticity. 
We emphasize that spreading is not crucial for wave propagation and that oscillations can also occur under confinement. However in contrast to our model, Ref.~\cite{deforet2014} recently proposed that oscillatory modes in confined layers can also be generated by stochastic motion of cells. Experimental tests that inhibit myosin based contractility or cell directionality can help discriminate between these different models.
%The model is amenable to generalizations to higher dimensions and can be adapted to study spreading in different planar geometries~\cite{SMprl14}. 

\begin{acknowledgments}
We thank Jeffrey Fredberg, James Butler, Jacob Notbohm and Michael K\"opf for useful discussions. The work at  Syracuse University was supported by the National Science Foundation (NSF) awards DMR-1305184  and DGE-1068780. MCM also acknowledges support from the Simons Foundation and from NSF award PHY11-25915  at the KITP of the University of California, Santa Barbara, and thanks KITP for its hospitality during completion of some of this work. SB gratefully acknowledges support from Kadanoff-Rice fellowship through NSF Materials Research Science and Engineering Center at the University of Chicago.
\end{acknowledgments}

%\bibliography{ref}

\newpage
\pagebreak
\hspace{0.3 in}
\section*{Supplementary Material}
\setcounter{figure}{0} 
\renewcommand{\figurename}{\textbf{Supplementary Figure}}
\renewcommand{\tablename}{\textbf{Supplementary Table}}
\renewcommand{\theequation}{S.\arabic{equation}}
\paragraph{Volume conservation and kinetic constraints.}
We begin by considering an ideal wound-healing geometry where the dimensions of the cell layer are by the instantenous height $h(t)$ (in the z-direction), x-length $L(t)$, and a y-length $d$. We further consider a thin film, letting $h(t)\ll L(t) \ll d$, such that the changes in $d$ are negligible compared to $L$ and $h$ and assume that there is translational invariance along the y-direction. If the initial height of the cell layer is $h_0$ and the length is $L_0$, we define vertical and horizontal strains as, $\varepsilon_h(t)=h(t)/h_0 -1$, and $\varepsilon_L=L(t)/L_0-1$. Volume conservation implies, $h(t) L(t)=$ constant. This leads to the following (related) kinetic constraints,
\begin{subequations}
\begin{gather}
\frac{\dot{h}}{h}=-\frac{\dot{L}}{L} \;,\\
\varepsilon_h=-\frac{\varepsilon_L}{1+\varepsilon_L}\;.
\end{gather}
\end{subequations}
Thus the instantaneous height is given by, $h(t)=h_0/(1+\varepsilon_L(t))$, where $\varepsilon_L(t)=u(L/2,t)-u(-L/2,t)$, $u$ being the displacement field in the x-direction.

\paragraph{Steady-state solution.} The homogeneous steady-state of Eqs. (1-2) in the main text is given by $u=0$ and $c=c_0$. This is a quiescent steady-state when the cell layer does not spread. To determine the inhomogeneous steady-state describing an expanded cell monolayer, we seek solutions of Eqs. (1-2) by setting $\partial_t u=0$ and $\partial_t c =0$. To make analytical progress, we first consider a cell monolayer of length much larger than the length scale of variations in the polarization, i.e $L >> \lambda$. In this limit the spreading force can be described by a signum function, $\tanh (x/\lambda) \sim \text{sign} (x/L)$ for $\lambda << L$. Next we linearize the active stress by considering small deviation of $c$ from its rest-state $c_0$,  $\log (c/c_0)\sim (c-c_0)/c_0$. The steady-state solution for the local stress is given by,
\begin{equation}
\label{sigma_ss}
\sigma_{\infty}(x)=-\dfrac{f_0}{2h_{\infty}}\left(|x| -\dfrac{L_{\infty}}{2}\right),
\end{equation}
where $L_{\infty}$ and $h_{\infty}$ are respectively the length and the height of the cell monolayer  at $t\rightarrow \infty$, given by $L_{\infty}=L_0 (1 + \langle \varepsilon_{\infty} \rangle)$ and $h_{\infty}=h_0/(1 + \langle \varepsilon_{\infty} \rangle)$. The spatial average of the steady-state strain $\langle \varepsilon_{\infty} \rangle$ is given by,
\begin{equation}
\label{av_strain}
\langle \varepsilon_{\infty} \rangle = \dfrac{1}{L_{\infty}}\int^{L_{\infty/2}}_{-L_{\infty/2}} \mathrm{d}x \varepsilon_{\infty}(x)\;.
\end{equation}
Combining Eq. \eqref{sigma_ss} with the equation for $c_{\infty}(x)$ we obtain a second order ordinary differential equation in $c_{\infty}$ whose analytical solution is,
\begin{widetext}
\[
c_{\infty}(x) = \dfrac{\alpha \tau}{B_{\text{eff}}}\left[-\dfrac{f_0}{2 h_{\infty}}\left(|x|-\dfrac{ L_{\infty}}{2} + \xi \dfrac{\left( \mathrm{e}^{-L_{\infty}/2\xi}-1\right)\cosh(x/\xi)}{\sinh(L_{\infty}/2\xi)} + \xi\mathrm{e}^{-|x|/\xi}\right) +\beta + \dfrac{Bc_0}{\alpha \tau}\right] \label{eq:c_ss}\;,
\]
\end{widetext}
where $B_{\text{eff}}= B + \alpha \beta \tau/c_0$ is the effective elastic modulus renormalized by active contractility and $\xi =\sqrt{\frac{B D \tau}{B_{\text{eff}}}}$ is a characteristic length scale associated with the spatial variation of the active agents. The steady-state solution for the average strain is,
\begin{equation}
\label{eq:strain_ss}
\varepsilon_{\infty}(x) =\dfrac{1}{B}\left(\sigma_{\infty}-\dfrac{\beta}{c_0} c_{\infty} + \beta \right)\;.
\end{equation}
The condition for the cell monolayer to expand is given by $\langle \varepsilon_{\infty} \rangle >0$ which leads to a critical value for the force density, $f^c_0=8 \beta h_{\infty} B_{\text{eff}}/B L_{\infty}$ above which the cell monolayer spreads.

\paragraph{Linear stability analysis.} In this section we examine the linear stability of the homogeneous steady-state. We render our system of equations dimensionless by letting $x\rightarrow x/L_0$ and $t\rightarrow f_0 t/\Gamma L_0$. Then new dimensionless parameters are $\tilde{B}=B/f_0$, $\tilde{\beta}=\beta/f_0$, $\tilde{\alpha}=\alpha \Gamma L_0/f_0c_0$,  $\tilde{\tau} =f_0 \tau /\Gamma L_0$, $\tilde{h}_0=h_0/L_0$ and $\tilde{D}=D\Gamma/f_0 L_0$. In these units, stress is measured in the units of $f_0$ and the concentration of contractile agents are expressed in units of $c_0$. In the following we drop the tilde notation over the dimensionless parameters for simplicity. Letting $\varepsilon (x,t)= \delta \varepsilon(x,t)$ and $c(x,t)=1 + \delta c(x,t)$, the linearized equations for the strain and the concentration fields are given by,
\begin{subequations}
\begin{gather}
 \partial_t \delta \varepsilon = h_0\left( B \partial_x^2 \delta \varepsilon + \beta \partial_x^2 \delta c \right)\;, \label{eq:delta_strain} \\
\partial_t \delta c  + \partial_t \delta \varepsilon= -\dfrac{1}{\tau} \delta c  + \alpha \delta \varepsilon  + D \partial_x^2 \delta c\label{eq:delta_c}\;. 
\end{gather}
\end{subequations}
%{\color{red} If we add p(x,t) in Eq. \eqref{eq:delta_strain} then we are not studying the linear stability analysis of the \textit{homogeneous} steady state because in this case there is no spreading force.}
Taking two spatial derivatives in Eq.~\eqref{eq:delta_c} and substituting the expression for $\partial_x^2\delta c=[h_0^{-1}\partial_t \varepsilon -B\partial_x^2\varepsilon ]/\beta$ (obtained from Eq.~\ref{eq:delta_strain}), we obtain the effective dynamics of strain fluctuations,
\begin{equation}
\tau \Gamma \partial_t^2\delta\varepsilon + \Gamma \partial_t \delta\varepsilon=h_0\left(B_\text{eff} + \eta_\text{eff}\partial_t-\tau B D \partial_x^2\right)\partial_x^2\delta\varepsilon \label{eq:sfluc}\;,
\end{equation}
where $B_\text{eff}=B+\alpha\beta\tau/c_0$ and $\eta_\text{eff}=(B-\beta + D\Gamma/h_0)\tau$. The dynamics of strain fluctuations are isomorphic to the dynamics of an driven damped Kelvin-Voigt material, with a characteristic frequency of oscillation $\omega_0$ given as a function of the wavelength $\lambda_0$,
\begin{equation}
\omega_0 (\lambda_0)=\frac{2\pi}{\lambda_0}\sqrt{\frac{h_0}{\tau\Gamma}\left[B_\text{eff} + \tau\left(\frac{2\pi}{\lambda_0}\right)^2 B D\right]} \label{eq:freq}\;.
\end{equation}
%\begin{figure}
%\centering
%\includegraphics[width=0.5\columnwidth]{figures/timeperiod.pdf}
%\caption{(color online). Comparison of the time period oscillation determined from numerical solution to Eq. (1-2) (red squares) and determined from the analytical expression in Eq.~\eqref{eq:freq} (black solid circles). The numerical value for the time period is obtained by performing a fast fourier transform on the strain rate at the midline of the monolayer. All the parameter values are the same as in Fig. 1 of the main text, with the addition $\lambda_0=L_0/2$.}
%\label{fig:timeperiod}
%\end{figure}

We compare the analytical prediction for the time period, $2\pi/\omega_0$, with the time period determined numerically by performing a Fast Fourier Transform on the solution for the strain rate at the center of the cell monolayer. The two values are in good agreement as shown in Fig. 3A of the main text for $\lambda_0=L_0/2$.

Next we look for solutions in the form $\delta \varepsilon, \delta c \sim \mathrm{e}^{i\omega t + iqx}$. The two eigenvalues controlling the dynamics of fluctuations are given by the following dispersion relations,
\begin{equation}
\label{eq:eigenvalue}
i\omega_{\pm}(q)=-\dfrac{b(q)}{2} \pm \dfrac{1}{2}\sqrt{[b(q)]^2-\dfrac{4h_0}{\tau} \left[(1 + D\tau q^2)B+\alpha \beta\right]q^2}
\end{equation}
where
$b(q)= \dfrac{1}{\tau} + \left[ h_0\left(B-\beta \right) + D\right]q^2$. If the coupling to the concentration of contractile elements is neglected ($\beta =0$), $i\omega_{+}=-h_0Bq^2$ and $i\omega_{-}=-\frac{1}{\tau}$ (we neglect the diffusion constant $D$ for simplicity assuming that its contribution is small). In this case the elastic deformations are stable and diffuse through the cell layer and no oscillations are observed.  When the coupling to the concentration field is considered, we find a region in the parameter space, $\mathcal{I}_m[\omega_{\pm}]>0$, where the linear fluctuations are unstable. The system is purely diffusive when $\mathcal{R}_e[\omega_{\pm}]=0$ and $\mathcal{I}_m[\omega_{\pm}]<0$. The region in the parameter space defined by the complex values of the fourier modes, i.e $\mathcal{R}_e[\omega_{\pm}] \neq 0$, describe oscillatory solutions. Furthermore, these propagating waves are stable when $\mathcal{I}_m[\omega_{\pm}]<0$.

\paragraph{Mean-field model.} The mean-field model admits an expanded steady-state solution for the gel with strain $\varepsilon_\infty$, and an average concentration of contractile agents $\bar{c}_\infty$. The strain fluctuations decay with similar dynamics as in Eq.~\eqref{eq:sfluc}. With $\delta\varepsilon=(L-L_\infty)/L_0$, we get for the dynamics of $\delta\varepsilon$,
\begin{equation}\label{eq:sfluc2}
\tau \gamma \frac{d^2\delta\varepsilon}{dt^2} + (\gamma + h_\infty\eta^\text{MF}_\text{eff}) \frac{d\delta\varepsilon}{dt} + h_\infty B^\text{MF}_\text{eff}\delta\varepsilon=0\;,
\end{equation}
where, $B_\text{eff}^\text{MF}=B-B\varepsilon_\infty-\beta(c_\infty/c_0-1)+\alpha\beta\tau/c_0$ and $\eta^\text{MF}_\text{eff}=\tau(B-B\varepsilon_\infty-\beta c_\infty/c_0)$. The time-period for oscillations is given by, $\omega_0^\text{MF}=\sqrt{h_\infty B_\text{eff}^\text{MF}/\tau\gamma}$. Since the mean-field model neglects diffusion, the effective viscosity $\eta^\text{MF}_\text{eff}$ characterizing the dissipation of strain-rate is less than $\eta_\text{eff}$. We thus add an additive correction $\eta_D$ to $\eta^\text{MF}_\text{eff}$ in order to accurately estimate the numerical value for the effective viscosity in our numerical analyses. Thus the condition for oscillatory solutions is given by, $4\tau\gamma h_\infty B_\text{eff}^\text{MF}>[\gamma + h_\infty(\eta_\text{eft}^\text{MF}+\eta_D)]^2$.

\paragraph{Turnovers in contractility are essential for stress wave propagation.}
See Supplementary Fig.~\ref{fig:fig1}.
\begin{figure}
\centering
\includegraphics[width=\columnwidth]{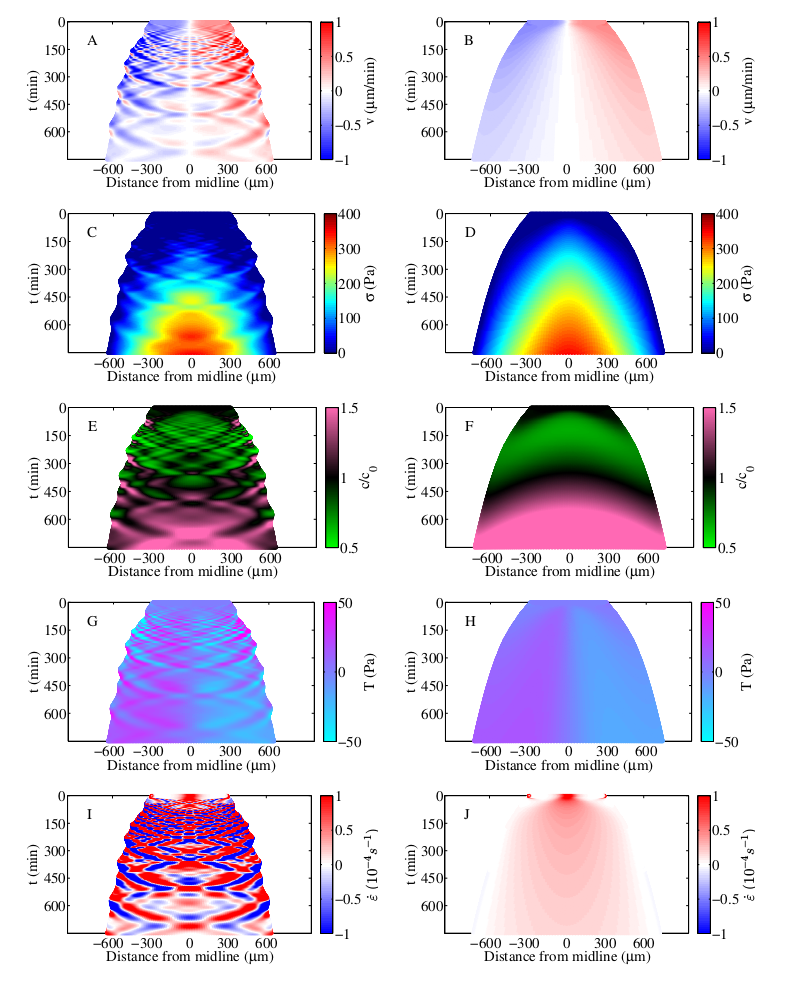}
\caption{(color online) Kymographs of (A,B) monolayer velocity $v(x,t)$, (C,D) monolayer stress $\sigma(x,t)$, (E,F) concentration of contractile elements, (G,H) traction stress $T(x,t)$ and (I,J) strain rate $\dot{\varepsilon}(x,t)$ in the non conserved case (left column) and conserved case (right column) of active units. Parameters: $B=120$ Pa, $\tau=350$ min, $(c_0/\alpha)=560$ $min$, $L_0=600$ $\mu$m, $h_0=6$ $\mu$m, $f_0=40$ Pa, $\Gamma=0.09$ nN min/$\mu$m$^3$, $D=26$ $\mu$m$^2$/min. For the magnitude of the contractile stress, we used $\beta=200$ Pa for the non conserved case and $\beta=30$ Pa for the conserved case.}
\label{fig:fig1}
\end{figure}

\paragraph{Complex oscillatory flows due to polarization fluctuations.} 
See Supplementary Fig.~\ref{fig:velo}.

\begin{figure}
\centering
\includegraphics[width=\columnwidth]{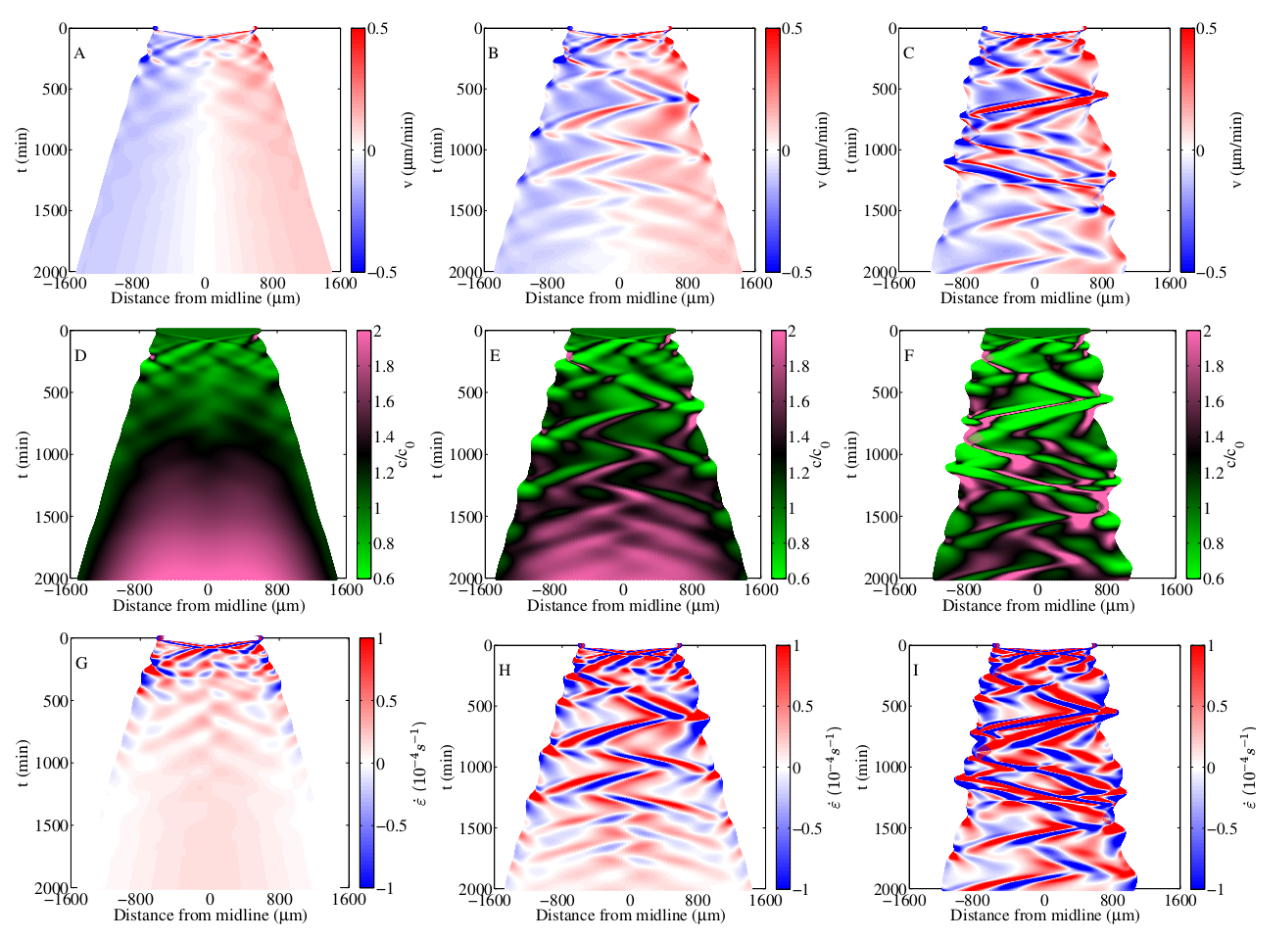}
\caption{(color online) Kymographs of local velocity (A-C), concentration of contractile elements (D-F) and strain rate (G-I) as the polarization induced tension $\beta'$ is increased (left to right). (A,D,G) X-waves, $\beta'=12$ nN/$\mu$m; (B,E,H) traveling stress pulse, $\beta'=17$ nN/$\mu$m; (C,F,I) complex oscillatory patterns, $\beta'=24$ nN/$\mu$m. Parameters: $w=4.3$ $\mu$m/min, $w'=0.21$ $\mu$m/min, $\kappa=193$ $\mu$m$^2$/min, $1/a=14$ min, $1/b=28$ min. Other parameter values are the same as in main text.}
\label{fig:velo}
\end{figure}

\paragraph{Planar dynamics of a spreading cell monolayer.}
In the main text, we describe an effective one-dimensional model for a spreading cell monolayer by assuming a translational invariance along y-direction. Here we formulate the dynamics of the spreading cell layer in two dimensions. The cell layer is modeled as an elastic medium whose local deformations are characterized by a displacement field $\textbf{u}(\textbf{r},t)$ about its initially undeformed state. Contractile activity in the constituent cells are described in terms of a scalar concentration field $c(\textbf{r},t)$ of actomyosin stress fibers. In addition, we describe local cell polarization by a vector field $\textbf{p}(\textbf{r},t)$, accounting for orientation of actin stress fibers. In the thin-film limit, where monolayer height is negligible compared to its planar dimensions, the thickness-averaged dynamics of the deformation field $\textbf{u}(\textbf{r},t)$ is given by,
\begin{align}
\label{Eq:deformation}
\Gamma \partial_t u_{i} = f_0 p_i + h(t) \partial_j \sigma_{ij}
\end{align}

where $\sigma_{ij}$ is the mechanical stress tensor given by a sum of elastic stress ($\sigma_{ij}^\text{el}$) and the active stresses ($\sigma_{ij}^a$) due to contractility and stress fiber polarization, $\sigma_{ij}=\sigma_{ij}^{\text{el}}+\sigma_{ij}^a$. Assuming linear elasticity, the elastic stress tensor is simply given by,
\begin{equation}
\sigma_{ij}^\text{el}=B \varepsilon_{kk} \delta_{ij}  + 2G\left(\varepsilon_{ij}-\dfrac{1}{2}\delta_{ij}\varepsilon_{kk} \right)\;,
\end{equation}
where $B$ and $G$ are the compressional and shear elastic moduli, respectively, and $\varepsilon_{ij}$ is the symmetrized strain tensor, $\varepsilon_{ij}=\dfrac{1}{2}\left(\partial_i u_j +\partial_ju_i \right)$.

The active stress comes from cell contractility, cell proliferation and polarization fluctuations. It is given by, 
\begin{align}
\sigma_{ij}^a=&-\Pi\ \delta_{ij} +\beta \log{\left(\dfrac{c}{c_0}\right)} \delta_{ij}+ \beta_1 \pmb{\nabla} \cdot \textbf{p}\ \delta_{ij} \nonumber \\ &+ \dfrac{\beta_2}{2}\left(\partial_i p_j + \partial_j p_i -\pmb{\nabla}\cdot{\bf p}\ \delta_{ij}\right)\;,
\end{align}
where $\Pi>0$ is an internal pressure due to cell proliferation, $\beta>0$ describes the mean contractile stress exerted by actomyosin units. The parameters $\beta_1$ and $\beta_2$ describe tensions induced by local gradients in the polarization field. 
The dynamics of the concentration field $c(\textbf{r},t)$ is given by,
\begin{align}
\label{Eq:concentration}
\partial_{t}c + \pmb{\nabla} \cdot (c \partial_t \textbf{u})=-\dfrac{1}{\tau} (c - c_0) + D \nabla^2 c + \alpha \varepsilon_{kk} +\alpha' \pmb{\nabla} \cdot \textbf{p}\;,
\end{align}
where $D$ is the diffusion constant; $\tau$ is the timescale of turnovers in contractility; $\alpha$ and $\alpha'$ are the rates of production of $c$ due to compressive mechanical and polarization strains.
The dynamics of the polarization field $\textbf{p}(\textbf{r},t)$ is given by,
\begin{align}
\label{Eq:polarization}
\partial_t p_i=&\left(a-b|\textbf{p}|^2 \right)p_i +\kappa \nabla^2 p_i - w_1 \partial_j \varepsilon_{ij} - w_2\partial_i \varepsilon_{kk} \nonumber \\ &+ w'\partial_i (c/c_0)\;,
\end{align}
where the first two terms (with $b>0$) allow for the onset of a homogeneously polar state, $\vert{\bf p}\vert=\sqrt{b/a}$, when $a>0$. Local cost of fluctuations in polarization is characterized by an isotropic stiffness $\kappa$; $w_1$, $w_2$ and $w'$ describe strength of alignment of cell polarization with gradients of elastic strain and the concentration field. Together, Eqs.~\eqref{Eq:deformation}, \eqref{Eq:concentration} and \eqref{Eq:polarization} describe the planar dynamics of the cell monolayer.

\paragraph{Choice of model parameters.}
While the model parameters are cell-type dependent, their values are chosen to quantitatively capture available experimental data in MDCK colonies~\cite{serra2012}. The model parameters are tuned to capture the experimental data on velocity, strain rates, traction stress and intercellular stresses, while the remaining values are chosen within the order of magnitudes reported in prior literature.  Specifically, we choose an initial length of the monolayer $L_0=600\ \mu$m, cross-sectional area $A=1000\ \mu {\rm m}^2$, spreading force $F_0=f_0 A=4\ {\rm nN}$ and friction $\gamma=\Gamma A=9\ {\rm nN~min}/\mu{\rm m}$. The timescale to reach a spread steady-state is thus $\tau_s=L_0\Gamma/f_0 \simeq 1400\ {\rm min}$. The timescale controlling activity turnover, $\tau$, and the timescale controlling the strain-concentration feedback, $c_0/\alpha$, are chosen smaller than $\tau_s$ but of the same order.
We assume that diffusion is negligible and tend to represent the tendency of neighboring cells to equalize the concentration of active agents. In our simulation we set the diffusion length to be $\sim 5\ \mu$m (one grid size) per minute.

%\bibliography{SIref}

\begin{thebibliography}{23}%
\makeatletter
\providecommand \@ifxundefined [1]{%
 \@ifx{#1\undefined}
}%
\providecommand \@ifnum [1]{%
 \ifnum #1\expandafter \@firstoftwo
 \else \expandafter \@secondoftwo
 \fi
}%
\providecommand \@ifx [1]{%
 \ifx #1\expandafter \@firstoftwo
 \else \expandafter \@secondoftwo
 \fi
}%
\providecommand \natexlab [1]{#1}%
\providecommand \enquote  [1]{``#1''}%
\providecommand \bibnamefont  [1]{#1}%
\providecommand \bibfnamefont [1]{#1}%
\providecommand \citenamefont [1]{#1}%
\providecommand \href@noop [0]{\@secondoftwo}%
\providecommand \href [0]{\begingroup \@sanitize@url \@href}%
\providecommand \@href[1]{\@@startlink{#1}\@@href}%
\providecommand \@@href[1]{\endgroup#1\@@endlink}%
\providecommand \@sanitize@url [0]{\catcode `\\12\catcode `\$12\catcode
  `\&12\catcode `\#12\catcode `\^12\catcode `\_12\catcode `\%12\relax}%
\providecommand \@@startlink[1]{}%
\providecommand \@@endlink[0]{}%
\providecommand \url  [0]{\begingroup\@sanitize@url \@url }%
\providecommand \@url [1]{\endgroup\@href {#1}{\urlprefix }}%
\providecommand \urlprefix  [0]{URL }%
\providecommand \Eprint [0]{\href }%
\providecommand \doibase [0]{http://dx.doi.org/}%
\providecommand \selectlanguage [0]{\@gobble}%
\providecommand \bibinfo  [0]{\@secondoftwo}%
\providecommand \bibfield  [0]{\@secondoftwo}%
\providecommand \translation [1]{[#1]}%
\providecommand \BibitemOpen [0]{}%
\providecommand \bibitemStop [0]{}%
\providecommand \bibitemNoStop [0]{.\EOS\space}%
\providecommand \EOS [0]{\spacefactor3000\relax}%
\providecommand \BibitemShut  [1]{\csname bibitem#1\endcsname}%
\let\auto@bib@innerbib\@empty
%</preamble>
\bibitem [{\citenamefont {Mammoto}\ and\ \citenamefont
  {Ingber}(2010)}]{mammoto2010}%
  \BibitemOpen
  \bibfield  {author} {\bibinfo {author} {\bibfnamefont {T.}~\bibnamefont
  {Mammoto}}\ and\ \bibinfo {author} {\bibfnamefont {D.~E.}\ \bibnamefont
  {Ingber}},\ }\href@noop {} {\bibfield  {journal} {\bibinfo  {journal}
  {Development}\ }\textbf {\bibinfo {volume} {137}},\ \bibinfo {pages} {1407}
  (\bibinfo {year} {2010})}\BibitemShut {NoStop}%
\bibitem [{\citenamefont {Keller}(2002)}]{keller2002}%
  \BibitemOpen
  \bibfield  {author} {\bibinfo {author} {\bibfnamefont {R.}~\bibnamefont
  {Keller}},\ }\href@noop {} {\bibfield  {journal} {\bibinfo  {journal}
  {Science}\ }\textbf {\bibinfo {volume} {298}},\ \bibinfo {pages} {1950}
  (\bibinfo {year} {2002})}\BibitemShut {NoStop}%
\bibitem [{\citenamefont {Fenteany}\ \emph {et~al.}(2000)\citenamefont
  {Fenteany}, \citenamefont {Janmey},\ and\ \citenamefont
  {Stossel}}]{fenteany2000}%
  \BibitemOpen
  \bibfield  {author} {\bibinfo {author} {\bibfnamefont {G.}~\bibnamefont
  {Fenteany}}, \bibinfo {author} {\bibfnamefont {P.~A.}\ \bibnamefont
  {Janmey}}, \ and\ \bibinfo {author} {\bibfnamefont {T.~P.}\ \bibnamefont
  {Stossel}},\ }\href@noop {} {\bibfield  {journal} {\bibinfo  {journal}
  {Current Biology}\ }\textbf {\bibinfo {volume} {10}},\ \bibinfo {pages} {831}
  (\bibinfo {year} {2000})}\BibitemShut {NoStop}%
\bibitem [{\citenamefont {Tsuji}\ \emph {et~al.}(2009)\citenamefont {Tsuji},
  \citenamefont {Ibaragi},\ and\ \citenamefont {Hu}}]{tsuji2009}%
  \BibitemOpen
  \bibfield  {author} {\bibinfo {author} {\bibfnamefont {T.}~\bibnamefont
  {Tsuji}}, \bibinfo {author} {\bibfnamefont {S.}~\bibnamefont {Ibaragi}}, \
  and\ \bibinfo {author} {\bibfnamefont {G.-f.}\ \bibnamefont {Hu}},\
  }\href@noop {} {\bibfield  {journal} {\bibinfo  {journal} {Cancer Research}\
  }\textbf {\bibinfo {volume} {69}},\ \bibinfo {pages} {7135} (\bibinfo {year}
  {2009})}\BibitemShut {NoStop}%
\bibitem [{\citenamefont {Weijer}(2009)}]{weijer2009}%
  \BibitemOpen
  \bibfield  {author} {\bibinfo {author} {\bibfnamefont {C.~J.}\ \bibnamefont
  {Weijer}},\ }\href@noop {} {\bibfield  {journal} {\bibinfo  {journal}
  {Journal of Cell Science}\ }\textbf {\bibinfo {volume} {122}},\ \bibinfo
  {pages} {3215} (\bibinfo {year} {2009})}\BibitemShut {NoStop}%
\bibitem [{\citenamefont {Manning}\ \emph {et~al.}(2010)\citenamefont
  {Manning}, \citenamefont {Foty}, \citenamefont {Steinberg},\ and\
  \citenamefont {Schoetz}}]{manning2010}%
  \BibitemOpen
  \bibfield  {author} {\bibinfo {author} {\bibfnamefont {M.~L.}\ \bibnamefont
  {Manning}}, \bibinfo {author} {\bibfnamefont {R.~A.}\ \bibnamefont {Foty}},
  \bibinfo {author} {\bibfnamefont {M.~S.}\ \bibnamefont {Steinberg}}, \ and\
  \bibinfo {author} {\bibfnamefont {E.-M.}\ \bibnamefont {Schoetz}},\
  }\href@noop {} {\bibfield  {journal} {\bibinfo  {journal} {Proceedings of the
  National Academy of Sciences}\ }\textbf {\bibinfo {volume} {107}},\ \bibinfo
  {pages} {12517} (\bibinfo {year} {2010})}\BibitemShut {NoStop}%
\bibitem [{\citenamefont {Mertz}\ \emph {et~al.}(2013)\citenamefont {Mertz},
  \citenamefont {Che}, \citenamefont {Banerjee}, \citenamefont {Goldstein},
  \citenamefont {Rosowski}, \citenamefont {Revilla}, \citenamefont {Niessen},
  \citenamefont {Marchetti}, \citenamefont {Dufresne},\ and\ \citenamefont
  {Horsley}}]{mertz2013}%
  \BibitemOpen
  \bibfield  {author} {\bibinfo {author} {\bibfnamefont {A.~F.}\ \bibnamefont
  {Mertz}}, \bibinfo {author} {\bibfnamefont {Y.}~\bibnamefont {Che}}, \bibinfo
  {author} {\bibfnamefont {S.}~\bibnamefont {Banerjee}}, \bibinfo {author}
  {\bibfnamefont {J.~M.}\ \bibnamefont {Goldstein}}, \bibinfo {author}
  {\bibfnamefont {K.~A.}\ \bibnamefont {Rosowski}}, \bibinfo {author}
  {\bibfnamefont {S.~F.}\ \bibnamefont {Revilla}}, \bibinfo {author}
  {\bibfnamefont {C.~M.}\ \bibnamefont {Niessen}}, \bibinfo {author}
  {\bibfnamefont {M.~C.}\ \bibnamefont {Marchetti}}, \bibinfo {author}
  {\bibfnamefont {E.~R.}\ \bibnamefont {Dufresne}}, \ and\ \bibinfo {author}
  {\bibfnamefont {V.}~\bibnamefont {Horsley}},\ }\href@noop {} {\bibfield
  {journal} {\bibinfo  {journal} {Proceedings of the National Academy of
  Sciences}\ }\textbf {\bibinfo {volume} {110}},\ \bibinfo {pages} {842}
  (\bibinfo {year} {2013})}\BibitemShut {NoStop}%
\bibitem [{\citenamefont {Serra-Picamal}\ \emph {et~al.}(2012)\citenamefont
  {Serra-Picamal}, \citenamefont {Conte}, \citenamefont {Vincent},
  \citenamefont {Anon}, \citenamefont {Tambe}, \citenamefont {Bazellieres},
  \citenamefont {Butler}, \citenamefont {Fredberg},\ and\ \citenamefont
  {Trepat}}]{serra2012}%
  \BibitemOpen
  \bibfield  {author} {\bibinfo {author} {\bibfnamefont {X.}~\bibnamefont
  {Serra-Picamal}}, \bibinfo {author} {\bibfnamefont {V.}~\bibnamefont
  {Conte}}, \bibinfo {author} {\bibfnamefont {R.}~\bibnamefont {Vincent}},
  \bibinfo {author} {\bibfnamefont {E.}~\bibnamefont {Anon}}, \bibinfo {author}
  {\bibfnamefont {D.~T.}\ \bibnamefont {Tambe}}, \bibinfo {author}
  {\bibfnamefont {E.}~\bibnamefont {Bazellieres}}, \bibinfo {author}
  {\bibfnamefont {J.~P.}\ \bibnamefont {Butler}}, \bibinfo {author}
  {\bibfnamefont {J.~J.}\ \bibnamefont {Fredberg}}, \ and\ \bibinfo {author}
  {\bibfnamefont {X.}~\bibnamefont {Trepat}},\ }\href@noop {} {\bibfield
  {journal} {\bibinfo  {journal} {Nature Physics}\ }\textbf {\bibinfo {volume}
  {8}},\ \bibinfo {pages} {628} (\bibinfo {year} {2012})}\BibitemShut {NoStop}%
\bibitem [{\citenamefont {Kruse}\ \emph {et~al.}(2005)\citenamefont {Kruse},
  \citenamefont {Joanny}, \citenamefont {J{\"u}licher}, \citenamefont {Prost},\
  and\ \citenamefont {Sekimoto}}]{kruse2005}%
  \BibitemOpen
  \bibfield  {author} {\bibinfo {author} {\bibfnamefont {K.}~\bibnamefont
  {Kruse}}, \bibinfo {author} {\bibfnamefont {J.-F.}\ \bibnamefont {Joanny}},
  \bibinfo {author} {\bibfnamefont {F.}~\bibnamefont {J{\"u}licher}}, \bibinfo
  {author} {\bibfnamefont {J.}~\bibnamefont {Prost}}, \ and\ \bibinfo {author}
  {\bibfnamefont {K.}~\bibnamefont {Sekimoto}},\ }\href@noop {} {\bibfield
  {journal} {\bibinfo  {journal} {The European Physical Journal E: Soft Matter
  and Biological Physics}\ }\textbf {\bibinfo {volume} {16}},\ \bibinfo {pages}
  {5} (\bibinfo {year} {2005})}\BibitemShut {NoStop}%
\bibitem [{\citenamefont {Bois}\ \emph {et~al.}(2011)\citenamefont {Bois},
  \citenamefont {J{\"u}licher},\ and\ \citenamefont {Grill}}]{bois2011}%
  \BibitemOpen
  \bibfield  {author} {\bibinfo {author} {\bibfnamefont {J.~S.}\ \bibnamefont
  {Bois}}, \bibinfo {author} {\bibfnamefont {F.}~\bibnamefont {J{\"u}licher}},
  \ and\ \bibinfo {author} {\bibfnamefont {S.~W.}\ \bibnamefont {Grill}},\
  }\href@noop {} {\bibfield  {journal} {\bibinfo  {journal} {Physical Review
  Letters}\ }\textbf {\bibinfo {volume} {106}},\ \bibinfo {pages} {028103}
  (\bibinfo {year} {2011})}\BibitemShut {NoStop}%
\bibitem [{\citenamefont {Marchetti}\ \emph {et~al.}(2013)\citenamefont
  {Marchetti}, \citenamefont {Joanny}, \citenamefont {Ramaswamy}, \citenamefont
  {Liverpool}, \citenamefont {Prost}, \citenamefont {Rao},\ and\ \citenamefont
  {Simha}}]{marchetti2013}%
  \BibitemOpen
  \bibfield  {author} {\bibinfo {author} {\bibfnamefont {M.}~\bibnamefont
  {Marchetti}}, \bibinfo {author} {\bibfnamefont {J.}~\bibnamefont {Joanny}},
  \bibinfo {author} {\bibfnamefont {S.}~\bibnamefont {Ramaswamy}}, \bibinfo
  {author} {\bibfnamefont {T.}~\bibnamefont {Liverpool}}, \bibinfo {author}
  {\bibfnamefont {J.}~\bibnamefont {Prost}}, \bibinfo {author} {\bibfnamefont
  {M.}~\bibnamefont {Rao}}, \ and\ \bibinfo {author} {\bibfnamefont {R.~A.}\
  \bibnamefont {Simha}},\ }\href@noop {} {\bibfield  {journal} {\bibinfo
  {journal} {Reviews of Modern Physics}\ }\textbf {\bibinfo {volume} {85}},\
  \bibinfo {pages} {1143} (\bibinfo {year} {2013})}\BibitemShut {NoStop}%
\bibitem [{\citenamefont {Kumar}\ \emph {et~al.}(2014)\citenamefont {Kumar},
  \citenamefont {Bois}, \citenamefont {J{\"u}licher},\ and\ \citenamefont
  {Grill}}]{kumar2014}%
  \BibitemOpen
  \bibfield  {author} {\bibinfo {author} {\bibfnamefont {K.~V.}\ \bibnamefont
  {Kumar}}, \bibinfo {author} {\bibfnamefont {J.~S.}\ \bibnamefont {Bois}},
  \bibinfo {author} {\bibfnamefont {F.}~\bibnamefont {J{\"u}licher}}, \ and\
  \bibinfo {author} {\bibfnamefont {S.~W.}\ \bibnamefont {Grill}},\ }\href@noop
  {} {\bibfield  {journal} {\bibinfo  {journal} {Physical Review Letters}\
  }\textbf {\bibinfo {volume} {112}},\ \bibinfo {pages} {208101} (\bibinfo
  {year} {2014})}\BibitemShut {NoStop}%
\bibitem [{\citenamefont {Arciero}\ \emph {et~al.}(2011)\citenamefont
  {Arciero}, \citenamefont {Mi}, \citenamefont {Branca}, \citenamefont
  {Hackam},\ and\ \citenamefont {Swigon}}]{arciero2011}%
  \BibitemOpen
  \bibfield  {author} {\bibinfo {author} {\bibfnamefont {J.~C.}\ \bibnamefont
  {Arciero}}, \bibinfo {author} {\bibfnamefont {Q.}~\bibnamefont {Mi}},
  \bibinfo {author} {\bibfnamefont {M.~F.}\ \bibnamefont {Branca}}, \bibinfo
  {author} {\bibfnamefont {D.~J.}\ \bibnamefont {Hackam}}, \ and\ \bibinfo
  {author} {\bibfnamefont {D.}~\bibnamefont {Swigon}},\ }\href@noop {}
  {\bibfield  {journal} {\bibinfo  {journal} {Biophysical Journal}\ }\textbf
  {\bibinfo {volume} {100}},\ \bibinfo {pages} {535} (\bibinfo {year}
  {2011})}\BibitemShut {NoStop}%
\bibitem [{\citenamefont {K{\"o}pf}\ and\ \citenamefont
  {Pismen}(2013)}]{kopf2013}%
  \BibitemOpen
  \bibfield  {author} {\bibinfo {author} {\bibfnamefont {M.~H.}\ \bibnamefont
  {K{\"o}pf}}\ and\ \bibinfo {author} {\bibfnamefont {L.~M.}\ \bibnamefont
  {Pismen}},\ }\href@noop {} {\bibfield  {journal} {\bibinfo  {journal} {Soft
  Matter}\ }\textbf {\bibinfo {volume} {9}},\ \bibinfo {pages} {3727} (\bibinfo
  {year} {2013})}\BibitemShut {NoStop}%
\bibitem [{\citenamefont {Banerjee}\ and\ \citenamefont
  {Marchetti}(2011)}]{banerjee2011}%
  \BibitemOpen
  \bibfield  {author} {\bibinfo {author} {\bibfnamefont {S.}~\bibnamefont
  {Banerjee}}\ and\ \bibinfo {author} {\bibfnamefont {M.~C.}\ \bibnamefont
  {Marchetti}},\ }\href@noop {} {\bibfield  {journal} {\bibinfo  {journal}
  {Soft Matter}\ }\textbf {\bibinfo {volume} {7}},\ \bibinfo {pages} {463}
  (\bibinfo {year} {2011})}\BibitemShut {NoStop}%
\bibitem [{\citenamefont {Banerjee}\ \emph {et~al.}(2011)\citenamefont
  {Banerjee}, \citenamefont {Liverpool},\ and\ \citenamefont
  {Marchetti}}]{banerjee2011a}%
  \BibitemOpen
  \bibfield  {author} {\bibinfo {author} {\bibfnamefont {S.}~\bibnamefont
  {Banerjee}}, \bibinfo {author} {\bibfnamefont {T.~B.}\ \bibnamefont
  {Liverpool}}, \ and\ \bibinfo {author} {\bibfnamefont {M.~C.}\ \bibnamefont
  {Marchetti}},\ }\href@noop {} {\bibfield  {journal} {\bibinfo  {journal} {EPL
  (Europhysics Letters)}\ }\textbf {\bibinfo {volume} {96}},\ \bibinfo {pages}
  {58004} (\bibinfo {year} {2011})}\BibitemShut {NoStop}%
\bibitem [{\citenamefont {Radszuweit}\ \emph {et~al.}(2013)\citenamefont
  {Radszuweit}, \citenamefont {Alonso}, \citenamefont {Engel},\ and\
  \citenamefont {B{\"a}r}}]{radszuweit2013}%
  \BibitemOpen
  \bibfield  {author} {\bibinfo {author} {\bibfnamefont {M.}~\bibnamefont
  {Radszuweit}}, \bibinfo {author} {\bibfnamefont {S.}~\bibnamefont {Alonso}},
  \bibinfo {author} {\bibfnamefont {H.}~\bibnamefont {Engel}}, \ and\ \bibinfo
  {author} {\bibfnamefont {M.}~\bibnamefont {B{\"a}r}},\ }\href@noop {}
  {\bibfield  {journal} {\bibinfo  {journal} {Physical review letters}\
  }\textbf {\bibinfo {volume} {110}},\ \bibinfo {pages} {138102} (\bibinfo
  {year} {2013})}\BibitemShut {NoStop}%
\bibitem [{\citenamefont {Pla{\c{c}}ais}\ \emph {et~al.}(2009)\citenamefont
  {Pla{\c{c}}ais}, \citenamefont {Balland}, \citenamefont {Gu{\'e}rin},
  \citenamefont {Joanny},\ and\ \citenamefont {Martin}}]{placcais2009}%
  \BibitemOpen
  \bibfield  {author} {\bibinfo {author} {\bibfnamefont {P.-Y.}\ \bibnamefont
  {Pla{\c{c}}ais}}, \bibinfo {author} {\bibfnamefont {M.}~\bibnamefont
  {Balland}}, \bibinfo {author} {\bibfnamefont {T.}~\bibnamefont {Gu{\'e}rin}},
  \bibinfo {author} {\bibfnamefont {J.-F.}\ \bibnamefont {Joanny}}, \ and\
  \bibinfo {author} {\bibfnamefont {P.}~\bibnamefont {Martin}},\ }\href@noop {}
  {\bibfield  {journal} {\bibinfo  {journal} {Physical review letters}\
  }\textbf {\bibinfo {volume} {103}},\ \bibinfo {pages} {158102} (\bibinfo
  {year} {2009})}\BibitemShut {NoStop}%
\bibitem [{\citenamefont {Dierkes}\ \emph {et~al.}(2014)\citenamefont
  {Dierkes}, \citenamefont {Sumi}, \citenamefont {Solon},\ and\ \citenamefont
  {Salbreux}}]{dierkes2014}%
  \BibitemOpen
  \bibfield  {author} {\bibinfo {author} {\bibfnamefont {K.}~\bibnamefont
  {Dierkes}}, \bibinfo {author} {\bibfnamefont {A.}~\bibnamefont {Sumi}},
  \bibinfo {author} {\bibfnamefont {J.}~\bibnamefont {Solon}}, \ and\ \bibinfo
  {author} {\bibfnamefont {G.}~\bibnamefont {Salbreux}},\ }\href@noop {}
  {\bibfield  {journal} {\bibinfo  {journal} {Physical review letters}\
  }\textbf {\bibinfo {volume} {113}},\ \bibinfo {pages} {148102} (\bibinfo
  {year} {2014})}\BibitemShut {NoStop}%
\bibitem [{\citenamefont {Kim}\ \emph {et~al.}(2013)\citenamefont {Kim},
  \citenamefont {Serra-Picamal}, \citenamefont {Tambe}, \citenamefont {Zhou},
  \citenamefont {Park}, \citenamefont {Sadati}, \citenamefont {Park},
  \citenamefont {Krishnan}, \citenamefont {Gweon}, \citenamefont {Millet} \emph
  {et~al.}}]{kim2013}%
  \BibitemOpen
  \bibfield  {author} {\bibinfo {author} {\bibfnamefont {J.~H.}\ \bibnamefont
  {Kim}}, \bibinfo {author} {\bibfnamefont {X.}~\bibnamefont {Serra-Picamal}},
  \bibinfo {author} {\bibfnamefont {D.~T.}\ \bibnamefont {Tambe}}, \bibinfo
  {author} {\bibfnamefont {E.~H.}\ \bibnamefont {Zhou}}, \bibinfo {author}
  {\bibfnamefont {C.~Y.}\ \bibnamefont {Park}}, \bibinfo {author}
  {\bibfnamefont {M.}~\bibnamefont {Sadati}}, \bibinfo {author} {\bibfnamefont
  {J.-A.}\ \bibnamefont {Park}}, \bibinfo {author} {\bibfnamefont
  {R.}~\bibnamefont {Krishnan}}, \bibinfo {author} {\bibfnamefont
  {B.}~\bibnamefont {Gweon}}, \bibinfo {author} {\bibfnamefont
  {E.}~\bibnamefont {Millet}},  \emph {et~al.},\ }\href@noop {} {\bibfield
  {journal} {\bibinfo  {journal} {Nature materials}\ }\textbf {\bibinfo
  {volume} {12}},\ \bibinfo {pages} {856} (\bibinfo {year} {2013})}\BibitemShut
  {NoStop}%
\bibitem [{SMp()}]{SMprl14}%
  \BibitemOpen
  \href@noop {} {}\bibinfo {note} {See Supplemental Material at [URL will be
  inserted by publisher] for details on the model, linear stability analysis
  and additional figures}\BibitemShut {NoStop}%
\bibitem [{\citenamefont {Trepat}\ \emph {et~al.}(2009)\citenamefont {Trepat},
  \citenamefont {Wasserman}, \citenamefont {Angelini}, \citenamefont {Millet},
  \citenamefont {Weitz}, \citenamefont {Butler},\ and\ \citenamefont
  {Fredberg}}]{trepat2009}%
  \BibitemOpen
  \bibfield  {author} {\bibinfo {author} {\bibfnamefont {X.}~\bibnamefont
  {Trepat}}, \bibinfo {author} {\bibfnamefont {M.~R.}\ \bibnamefont
  {Wasserman}}, \bibinfo {author} {\bibfnamefont {T.~E.}\ \bibnamefont
  {Angelini}}, \bibinfo {author} {\bibfnamefont {E.}~\bibnamefont {Millet}},
  \bibinfo {author} {\bibfnamefont {D.~A.}\ \bibnamefont {Weitz}}, \bibinfo
  {author} {\bibfnamefont {J.~P.}\ \bibnamefont {Butler}}, \ and\ \bibinfo
  {author} {\bibfnamefont {J.~J.}\ \bibnamefont {Fredberg}},\ }\href@noop {}
  {\bibfield  {journal} {\bibinfo  {journal} {Nature Physics}\ }\textbf
  {\bibinfo {volume} {5}},\ \bibinfo {pages} {426} (\bibinfo {year}
  {2009})}\BibitemShut {NoStop}%
\bibitem [{\citenamefont {Deforet}\ \emph {et~al.}(2014)\citenamefont
  {Deforet}, \citenamefont {Hakim}, \citenamefont {Yevick}, \citenamefont
  {Duclos},\ and\ \citenamefont {Silberzan}}]{deforet2014}%
  \BibitemOpen
  \bibfield  {author} {\bibinfo {author} {\bibfnamefont {M.}~\bibnamefont
  {Deforet}}, \bibinfo {author} {\bibfnamefont {V.}~\bibnamefont {Hakim}},
  \bibinfo {author} {\bibfnamefont {H.}~\bibnamefont {Yevick}}, \bibinfo
  {author} {\bibfnamefont {G.}~\bibnamefont {Duclos}}, \ and\ \bibinfo {author}
  {\bibfnamefont {P.}~\bibnamefont {Silberzan}},\ }\href@noop {} {\bibfield
  {journal} {\bibinfo  {journal} {Nature communications}\ }\textbf {\bibinfo
  {volume} {5}} (\bibinfo {year} {2014})}\BibitemShut {NoStop}%
\end{thebibliography}%

\end{document}